\documentclass[12pt,preprint]{aastex}
\title{IR Dust Bubbles II: Probing the Detailed Structure and Young Massive
  Stellar Populations of Galactic HII Regions} 
\author{C.
  Watson\altaffilmark{1}, T. Corn\altaffilmark{1}, 
  E.B. Churchwell\altaffilmark{2}, B.L. Babler\altaffilmark{2},  M.S. Povich\altaffilmark{2}, M.R.  Meade\altaffilmark{2},
  B.A.  Whitney\altaffilmark{3}}
\altaffiltext{1}{Manchester College, Dept. of Physics, 604 E. College Ave.,
  North Manchester, IN 46962} 
\altaffiltext{2}{Univ. of Wisconsin - Madison,
  Dept. of Astronomy, 475 N. Charter St., Madison, WI 53716}
\altaffiltext{3}{Space Science Institute, 4750 Walnut St. Suite 205, Boulder, CO 80301}
\begin{document}
\begin{abstract}
We present an analysis of late-O/early-B-powered, parsec-sized bubbles and associated star-formation using 2MASS, GLIMPSE, MIPSGAL and MAGPIS surveys. Three bubbles were selected from the Churchwell et al. (2007) catalog. We confirm that the structure identified in Watson et al. (2008) holds in less energetic bubbles, i.e. a PDR, identified by 8 $\mu$m emission due to PAHs surrounds hot dust, identified by 24 $\mu$m emission and ionized gas, identified by 20 cm continuum. We estimate the dynamical age of two bubbles by comparing bubble sizes to numerical models of Hosokawa \& Inutsuka (2006). We also identify and analyze candidate young stellar objects (YSOs) using SED fitting and identify sites of possible triggered star-formation. Lastly, we identify likely ionizing sources for two sources based on SED fitting.
\end{abstract}
\section{Introduction}
Massive stars strongly influence their surrounding environment throughout
their lifetime via stellar winds, ionizing radiation, heating of dust and expansion of their HII regions. Some of these processes may trigger
second-generation star-formation by compressing neighboring pre-existing molecular material to the point of
gravitational instability. Observing massive star-formation regions, however,
has been hampered by large UV and optical extinction. These regions are observed at IR, radio and x-ray wavelengths where extinction is significantly smaller.

Churchwell et al. (2006, 2007) analyzed mid-IR (MIR) images from the
Spitzer-GLIMPSE project, a survey of the Galactic plane ($|b|<$1$^\circ$,
$|l|<$65$^\circ$, Benjamin et al. 2003), and found bubbles of diffuse emission to be a signature
structure in the ISM at MIR wavelengths. They catalogued almost 600 bubbles (an admittedly incomplete catalogue) in
the GLIMPSE survey area. They argued based on the location and coincidence
with known HII regions that many of the MIR bubbles are produced by O and early-B stars. Watson et
al. (2007) analyzed the structure of three bubbles and associated
star-formation using surveys in the mid-IR (GLIMPSE and MIPSGAL) and radio continuum
(MAGPIS). They concluded that the general structure of the bubbles is a
photo-dissociated region (PDR), visible in the 5.8 and 8 $\mu$m IRAC bands on Spitzer,
which encloses ionized gas (observed at 20 cm) and hot dust (observed at 24
$\mu$m).  One bubble (N49) showed evidence of a cavity at 24 $\mu$m and 20 cm,
indicating that hot dust and ionized gas have been evacuated by stellar winds. They also
characterized the young stellar objects (YSOs) associated with each bubble and
identified sites of probable triggered star-formation and sources likely responsible for
ionizing hydrogen and exciting the PDR.

Deharveng and collaborators have studied PDRs and triggered star-formation
around the HII regions Sh 217 and Sh 219 (Deharveng et al. 2003a), Sh 104
(Deharveng et al. 2003b), RCW 79 (Zavagno et al. 2006), SH2-219 (Zavagno et al. 2006) and RCW 120 (Zavagno et
al. 2007). They identified several sites of probable triggered
star-formation, some by the collapse of pre-existing clouds (observed at 1.2
mm continuum) and some by the collect-and-collapse mechanism (see Whitworth, 1994 and Elmegreen,
1998 and references therein). Briefly, the collect-and-collapse mechanism posits that ambient ISM is swept-up by an expanding HII region, increasing in density until one or more subcomponents become gravitationally unstable and collapse, leading to star-formation. For some of these regions, they estimated ages
for the HII region and masses for the surrounding mm-clumps. They also classified
the YSOs in the regions into the standard classes
based on near and mid-IR colors.

Here, we analyze the gas and dust structure in three GLIMPSE-identified bubbles in the
Churchwell et al. (2007) catalog. We also measure star-formation activity and
identify YSOs to characterize possible triggering mechanisms. Three
sources were chosen for their range of sizes, association with likely triggered star formation, bubble dynamical ages
and spectral-types of the ionizing star(s). All are at low-longitudes ($|l|
<$10$^\circ$).  In \S2 we introduce each source and the surveys used. In \S3,
we discuss the relative position of gas and dust components (PAHs, ionized gas
and dust), identify YSOs and analyze their properties and identify candidate ionizing stars in each bubble.
In \S4, we discuss the results in the context of triggered star-formation
mechanisms.  Our main conclusions are summarized in \S5.

\section{Data}
Data were collected from four large-scale surveys: 2MASS, GLIMPSE, MIPSGAL and
MAGPIS. The 2MASS All-Sky Point-Source Catalog covers over 99\% of the sky at bands J, H and K$_s$. Along with the mosaiced images from GLIMPSE, we used the GLIMPSE Point
Source Catalog (GPSC), a 99.5\% reliable catalog of point sources observed in
the Spitzer-IRAC bands (3.6, 4.5, 5.8 and 8.0 $\mu$m). IRAC has a
resolution of 1.5'' to 1.9'' (3.6 to 8.0 $\mu$m).  See the GLIMPSE Data
Products
Description\footnote{http://www.astro.wisc.edu/glimpse/glimpse1\_dataprod\_v2.0.pdf}
for details. MIPSGAL is a survey of the same region as GLIMPSE, using the MIPS
instrument (24 $\mu$m and 70 $\mu$m) on Spitzer. MIPSGAL has a resolution of
6'' at 24 $\mu$m and 18'' at 70 $\mu$m. MAGPIS is a survey of the Galactic
plane at 20 cm and 6 cm using the VLA in configurations B, C and D combined
with the Effelsburg 100m single dish telescope (White, Becker \& Helfand 2005).  MAGPIS has a resolution of
6'' at both wavelengths.

We analyze CN138, CN108 and CS57 from the Churchwell et al. (2007) catalog of
bubbles. All sources are within 10$^\circ$ of the Galactic center and have not been studied in detail previously. CN138 is
near IRAS source 18073-2046, which has been studied by Mateen, Hofner \&
Araya (2006) in SO J=1-0, Scoville et al. (1987) in radio recombination lines, Walsh et al. (1997) in methanol masers and Slysh et al. (1999) in methanol masers. However, IRAS 18073-2046 appears to be coincident with CN139 in the bubble catalog, a large, complex bubble. Here, we choose to isolate our
analysis to the morphologically simpler and smaller CN138, which has not been studied previously. There are two IRAS
sources present toward CN108: IRAS 18028-2208 and IRAS 18029-2213. A
Wolf-Rayet star was identified by Shara et al. (1999) at $l$ = 8.02$^\circ$ $b$ =
-0.42$^\circ$ on the boundary of CN108 in projection. CN108 was also observed by Lockman et al. (1996) in the radio recombination lines H109$\alpha$ and H111$\alpha$ with the Green Bank 140-ft telescope. Near CS57, IRAS 17262-3435 is observed in the PDR shell and IRAS 17258-3432 is observed outside the PDR shell. No other known observations of these bubbles exist besides the surveys summarized above. Velocity measurements are available for all the bubbles and are summarized in \S 3.

\section{Results}
All three sources show the same basic structure of gas and dust that Watson et al. (2008) observed in 3 other bubbles: A PDR shell (identified by 5.8 $\mu$m and 8 $\mu$m PAH emission) surrounding hot dust (identified by 24 $\mu$m emission). In two sources (CN138 and CN108), the PDR also surrounds ionized gas (identified by radio continuum emission that overlaps with the 24 $\mu$m emission). We now present the mid-IR observations, YSO properties and ionizing star candidates for the three selected sources (CN138, CN108 and CS57).

\subsection{CN138}
\begin{figure}
\caption{Top: CN138 shown in 24 $\mu$m (red), 8 $\mu$m (green) and 4.5 $\mu$m (blue). 20 cm
  in contours at 1.5, 1.8, 2.1 and 2.4 mJy/Beam. The distance indicated at the lower-right is calculated using a kinematic distance of 4.3 kpc. Bottom: Slice at $l$=9.83$^\circ$. 20 cm (dotted, x 10$^6$), 24 $\mu$m (dashed) and 8.0
  $\mu$m (solid, x 5). The 20 cm and 24 $\mu$m are largely contained
  within the 8 $\mu$m peaks which represent the shell.}
\plotone{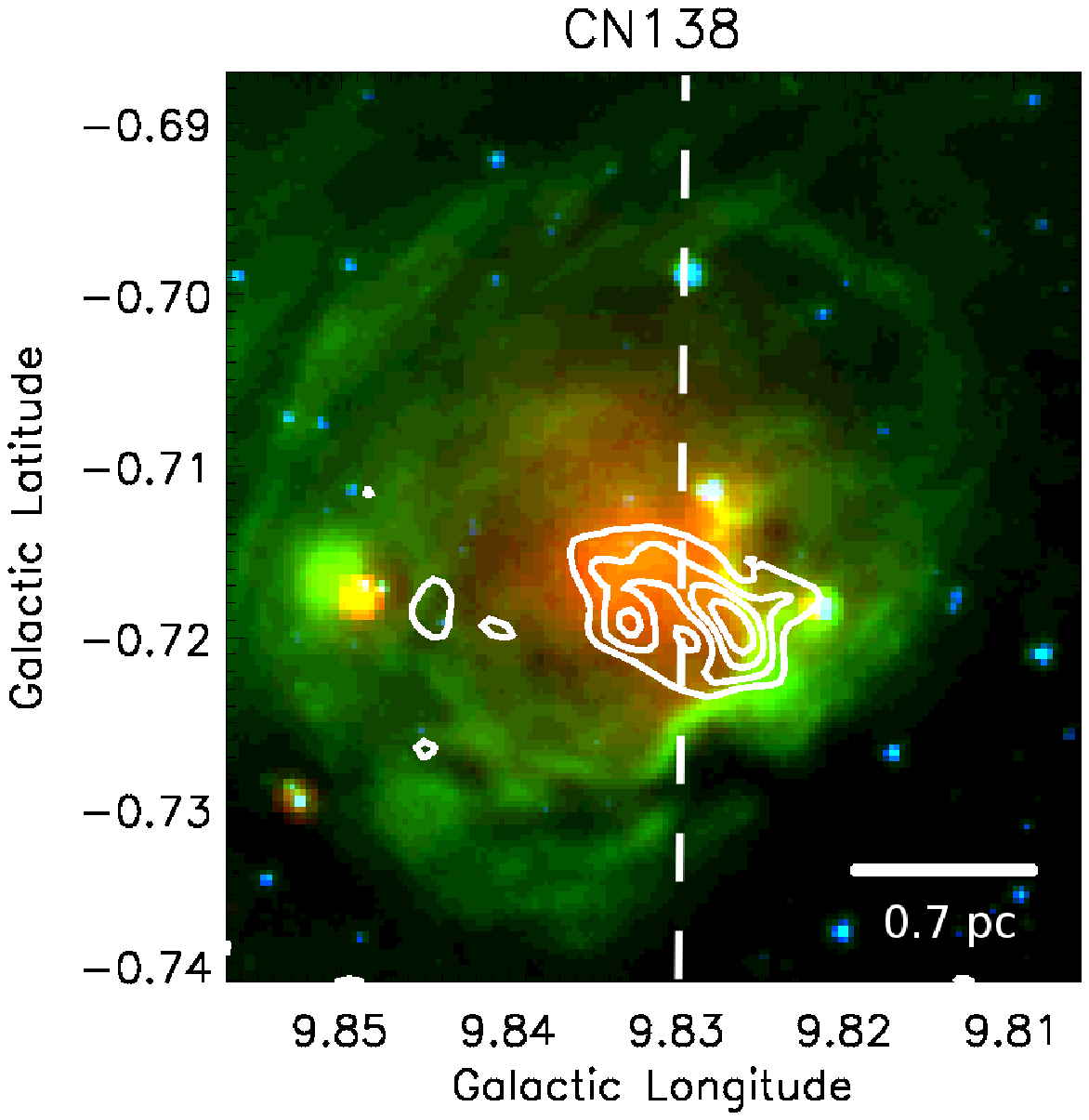}
\plotone{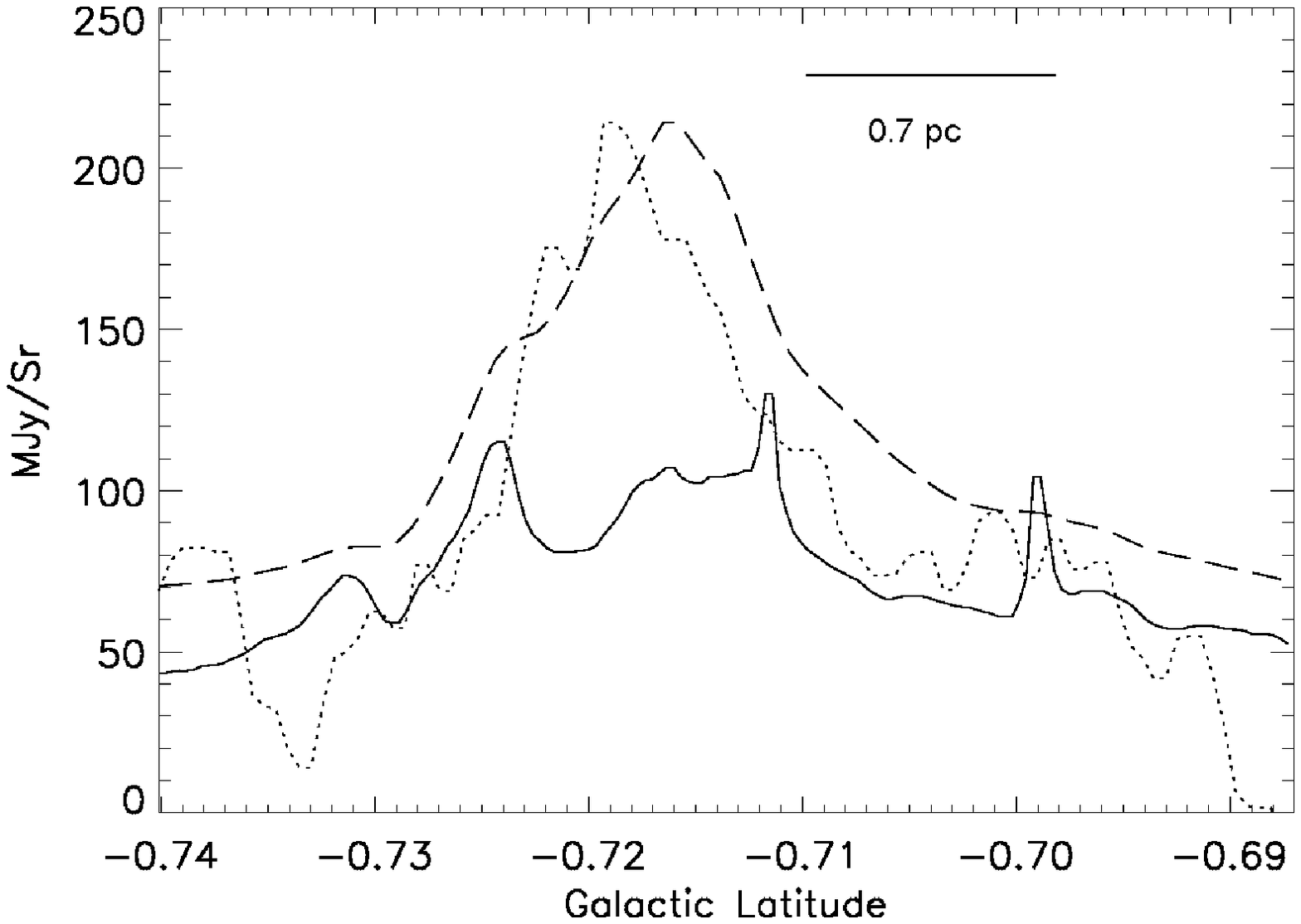}
\label{cn138_3color}
\end{figure}

CN138 has a shell morphology at 8 $\mu$m that surrounds 24 $\mu$m and 20
cm emission (see Fig \ref{cn138_3color}). Its kinematic distance is
4.3$\pm$0.6 kpc (based on a radio recombination line velocity of 34 km
s$^{-1}$, see Scoville et al, 1987, and the rotation model of Brand \& Blitz
1990). Errors are calculated assuming departures from circular velocities of 10 km s$^{-1}$. We measure the average radius to the inner boundary of the 8 $\mu$m shell to be
$\sim$80'' ($\sim$1.7 pc), the FWHM of the 24 $\mu$m emission to be $\sim$54'' ($\sim$1.1 pc) and the FWHM
at 20 cm emission to be 36'' (0.8 pc). The 24 $\mu$m and 20 cm difference in radii may be due to the low sensitivity of the 20 cm observations. As shown in Fig~\ref{cn138_3color} (bottom), the 20 cm and 24 $\mu$m emission peaks significantly overlap, while the 8 $\mu$m emission peaks are offset and surround the 20 cm and 24 $\mu$m emission peaks. The integrated flux density at 20 cm is
0.19 Jy, indicating an ionizing flux of N$_{Ly}$=2.9$\times$10$^{47}$ photons
s$^{-1}$, equivalent to a B0-B0.5 star (based on extrapolating Martins,
Schaerer \& Hillier 2005, hereafter MSH05, to B-type stars).

\begin{figure}
\caption{CN138: Candidate YSOs identified using the numerical models and SED
  fitting explained in Robitaille et al. (2007) overlaid on an 24 $\mu$m (red),
  8.0 $\mu$m (green) and 4.5 $\mu$m (blue) image. Stage I sources are shown in yellow, stage II in cyan and stage III in red. Two sites of possible triggered
  star-formation are visible to the left and right of the bubble center.
  CN139, a nearby and larger bubble, is present to the left of CN138.}
\label{cn138yso}
\includegraphics[width=5in]{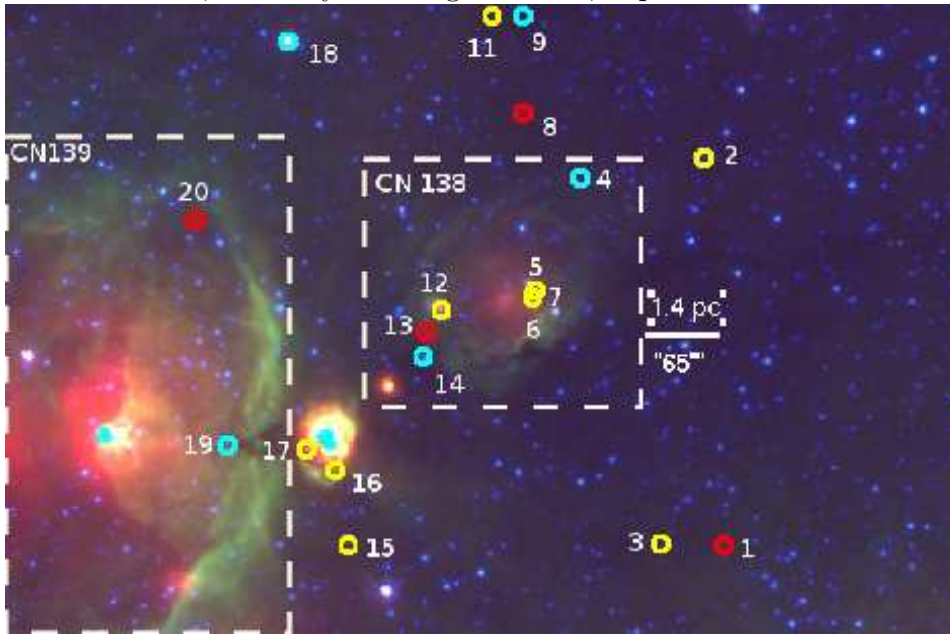}
\end{figure}

There are 1850 sources in the GPSC within 300'' (6.7 pc) of the center of CN138. We chose a relatively large area surrounding CN138 to show the star-formation associated with CN139. We
performed point source photometry on the MIPSGAL images and cross-correlated
the resultant 24 $\mu$m sources with corresponding GPSC sources. These sources
were then analyzed using the YSO-fitting method of Robitaille et al. (2007).
Briefly, this method involves a grid of Monte Carlo radiative transfer models of YSOs
with specified stellar masses, luminosities, disk masses, mass accretion rates and
line-of-sight inclinations (Robitaille et al. 2006). Observations from J-band to 24 $\mu$m are fit
using a $\chi^2$-minimization technique. The range of models that fit the
observations within the observational errors give an implied range of YSO
physical properties. All the YSOs surrounding CN138 are shown in
Fig.~\ref{cn138yso} and the range of stellar masses, total luminosities and envelope accretion rates are given in Table \ref{ysotable}.   The large area in Fig \ref{cn138yso} is shown to demonstrate CN139 (to the lower left) and the star-formation that is not associated with the shell of CN138 (see below). Each YSO is classified as stage 0 if $\dot M_{env}$/M$_* >$ 10$^{-6}$ yr$^{-1}$, stage I if $\dot M_{env}$/M$_* <$ 10$^{-6}$ yr$^{-1}$ and $\dot M_{disk}$/M$_* >$ 10$^{-6}$ yr$^{-1}$, or stage III if $\dot M_{disk}$/M$_* <$ 10$^{-6}$ yr$^{-1}$ and $\dot M_{env}$/M$_* <$ 10$^{-6}$ yr$^{-1}$, following Robitaille et al. (2006).

Some of these candidate YSOs are likely foreground or background YSOs unassociated with the bubble as Povich et al. (2008) found toward the M17 complex. By analyzing an off-source control sample using the SKY model of IR point sources, they concluded that a majority of the contaminates were YSOs at unknown distances. We expect that contamination of our sample with foreground or background sources would be higher at low galactic longitudes, since the line of sight covers more volume than the earlier analysis. For these sources, the properties given in Table \ref{ysotable} are incorrect.

There are two sites of possible triggered star-formation along the shell of
CN138, to the right and left of the bubble center. Both groups of YSOs are
low-to-intermediate mass (M$_*$ $<$ 10 M$_\odot$). The lack of YSOs and dimmer
8 $\mu$m PAH emission along the rest of the rim implies that the density of
gas may be higher to the east and west (assuming roughly equal illumination by
the hot star(s) that ionize CN138). This source
is qualitatively similar to the rim surrounding RCW 79 (Zavagno et al. 2006).
In their study of molecular gas and GLIMPSE data observed toward RCW 79, Zavagno et al.  (2006)
pointed out that the YSOs along the edge of the HII region formed distinct groups. One prediction of the collect-and-collapse model is that YSOs will form in such groups along the shock front of the expanding HII region. The groups are related to the gravitational instability length-scale, modified by effects due to the expanding shock wave, pressure external to the shock wave and the shock layer thickness. Zavagno et al. (2006) concluded that this mechanism was operative in RCW 79. In CN138 it also appears that the YSOs are formed into two distinct groups.

\begin{figure}
\caption{CN138: Candidate ionizing stars identified using the method of Watson
  et al. (2007) overlaid on an 8 $\mu$m (red), 4.5 $\mu$m (green) and 3.6
  $\mu$m (blue) image. See Table \ref{obtable} for star properties.}
\label{obimage}
\includegraphics[angle=0,width=6in]{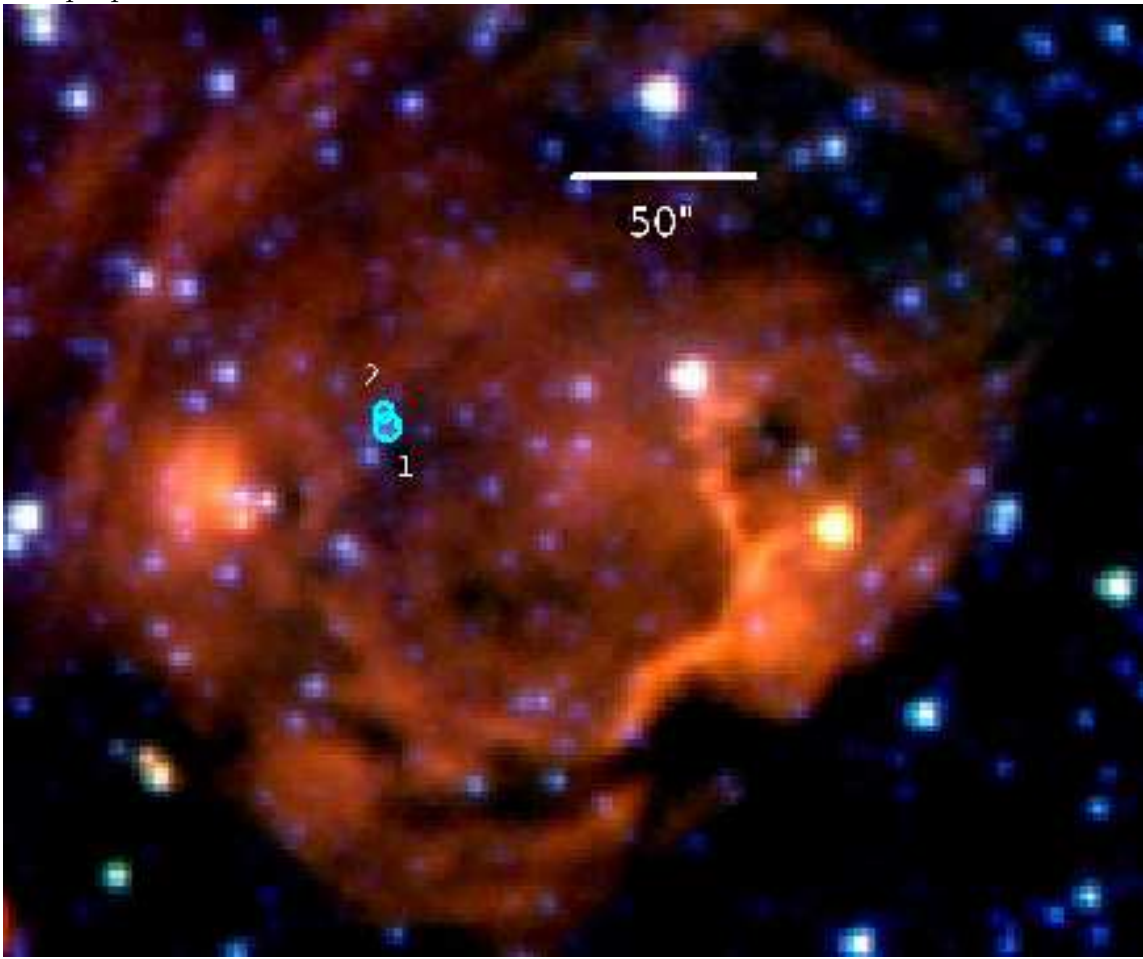}
\end{figure}

GPSC+MIPSGAL sources have been analyzed to find the ionizing star(s)
responsible for CN138 following the process outlined in Watson et al. (2008). Since we lack spectra of these stars, this method was developed to identify candidate ionizing sources.
Briefly, we identify those sources whose SEDs are consistent with an early-B
or O-type star at the distance of the CN138 bubble with the following constraints: the source must be fit by a hot stellar
photosphere model with no circumstellar emission, at the distance of the bubble (4.3
kpc, see above), there must be some extinction and the source must lie inside projected bubble boundaries. Two sources were found consistent with the above criteria.
Their locations are shown in Fig~\ref{obimage} and properties are given in
Table~\ref{obtable}. These sources are two sub-classes earlier than implied by the radio
continuum emission. One of these sources may be the ionizing star but the other is likely a foreground, cooler star. The discrepancy between estimated spectral type of the ionizing star from the observed radio continuum is probably primarily due to dust absorption of UV photons in the HII region, which are not counted by radio continuum emission. There are, of course, uncertainties in determining spectral types using this method as well. Both of these possible ionizing stars are significantly off-center in projection with respect to the bubble. Since the bubble is not circular, however, it may be reasonable for the ionizing source to not be perfectly centered. However, the 24 $\mu$m emission peak is significantly offset from the candidate ionizing stars. These offsets may indicate that we have not identified the ionizing source for this bubble. 

\subsection{CN108}
\begin{figure}
\caption{Top: CN108 shown in 24 $\mu$m (red), 8 $\mu$m (green) and 4.5 $\mu$m (blue). 20 cm
  contoured at 1.1 mJy. Note that the 20 cm is probably over-resolved and thus missing flux. Bottom: The slice is at l=8.1$^\circ$ showing 20 cm (dotted, x 10$^6$), 24 $\mu$m (dashed), and 8.0 $\mu$m (solid, x 5). The 24 $\mu$m and 20 cm emission
  is concentrated between the 8 $\mu$m peaks that indicate the shell. The 8
  $\mu$m spike at $b$ = -0.53 is caused by a star.}
\label{cn108_3color}
\includegraphics[width=5in]{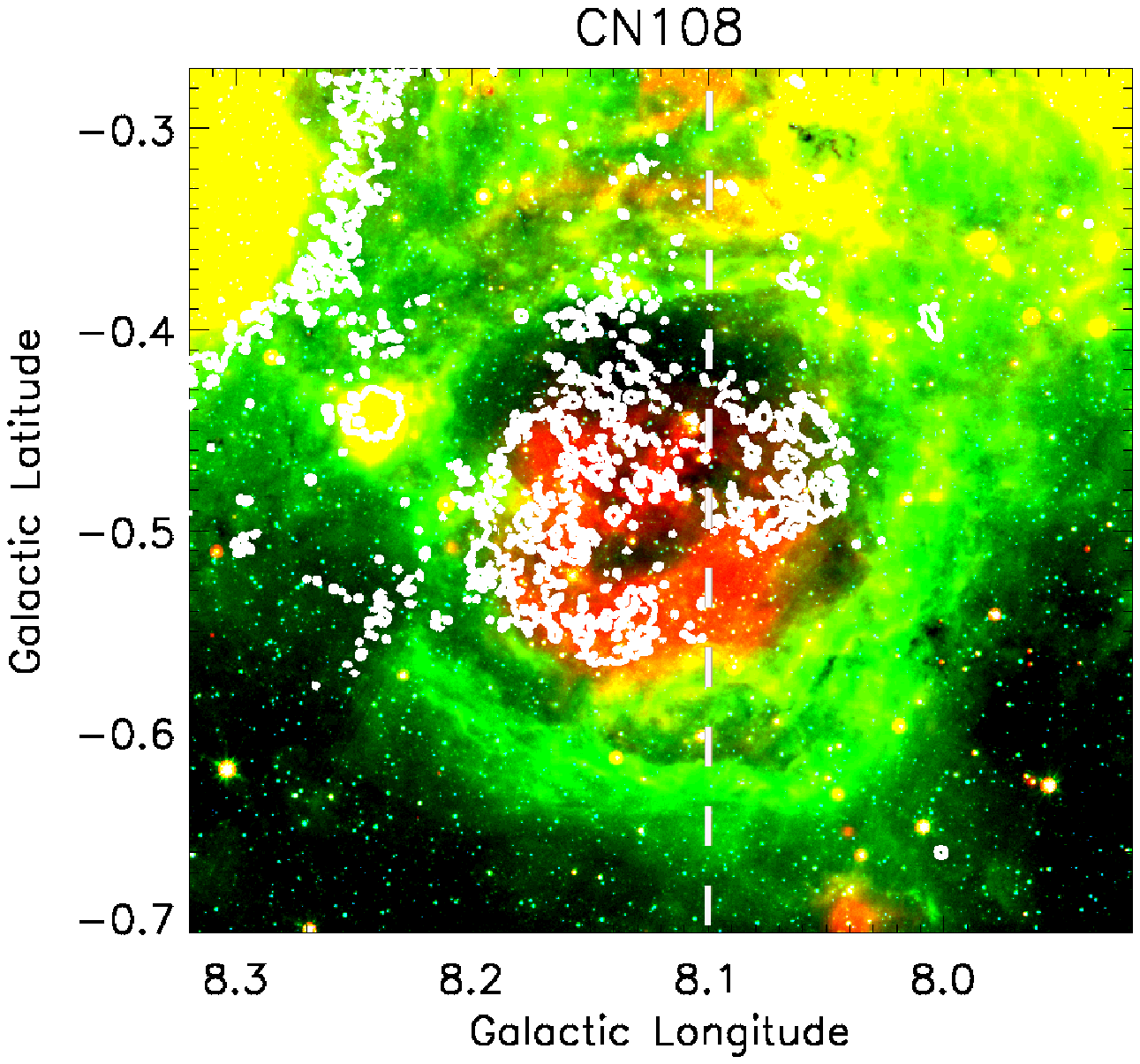}
\includegraphics[width=5in]{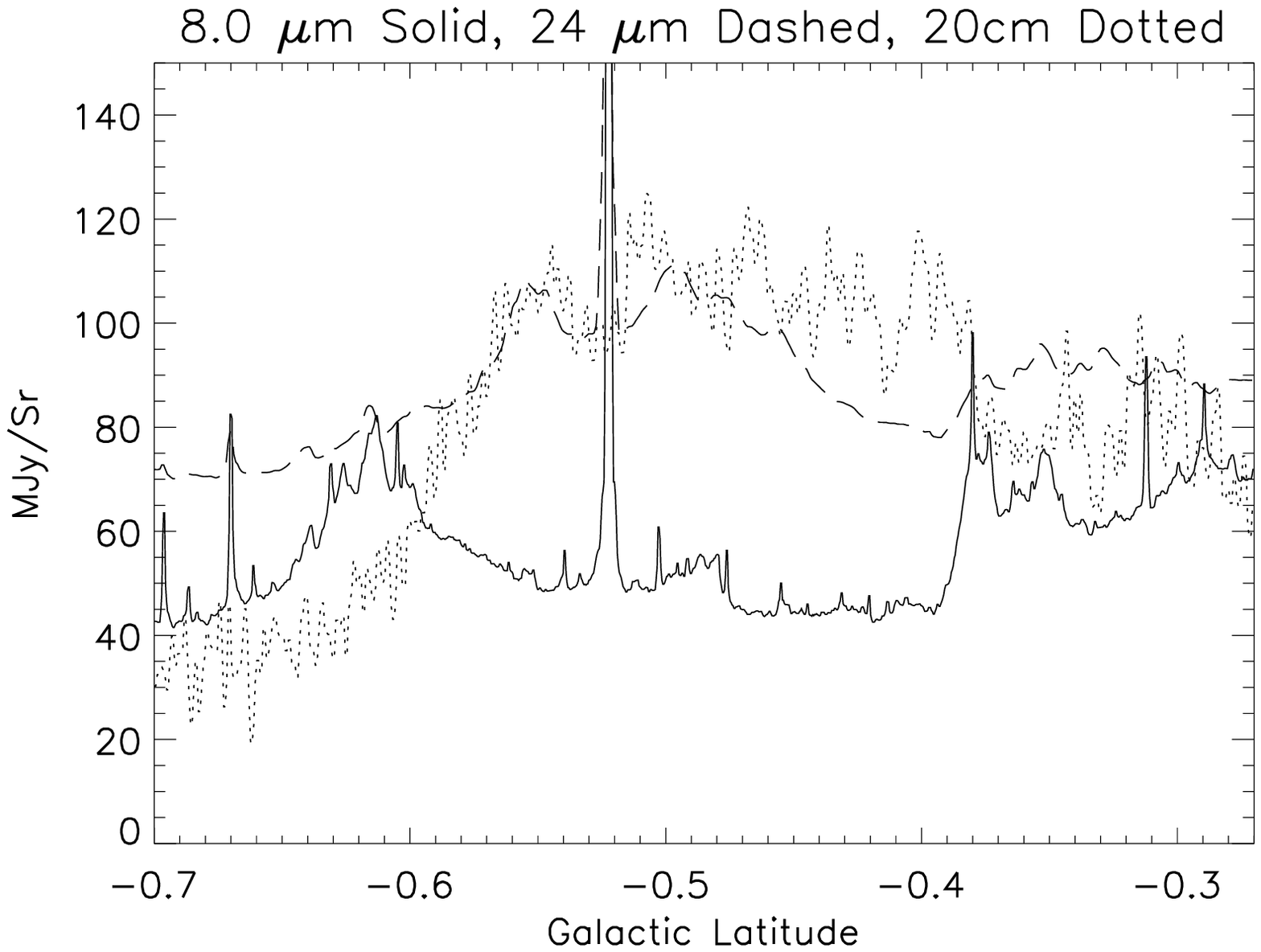}
\end{figure}

CN108 has a shell morphology at 8 $\mu$m that surrounds a smaller shell of
24 $\mu$m and 20 cm emission (see Fig. \ref{cn108_3color}). Its kinematic
distance is 4.9 kpc (based on a radio recombination line velocity measurement
of 37 km s$^{-1}$ by Lockman, Pisano \& Howard 1996). We measure the average inner radius of the 8 $\mu$m shell to be 340'' (8.0 pc) and the average outer
radius to be 520'' (12 pc). The integrated 20 cm flux density is 13.6 Jy,
equivalent to an ionizing flux of 1.9 $\times$ 10$^{48}$ photons s$^{-1}$ or
a single O8V star (MSH05). The radio continuum, however, appears strongly over-resolved and this flux density is likely an underestimate.

\begin{figure}
\caption{CN108: Candidate YSOs overlaid on an 24 $\mu$m (red),
  8.0 $\mu$m (green) and 4.5 $\mu$m (blue) image. Stage I sources are shown in yellow, stage II
  in cyan and stage III in red.}
\label{cn108ysomap}
\includegraphics[angle=0,width=6in]{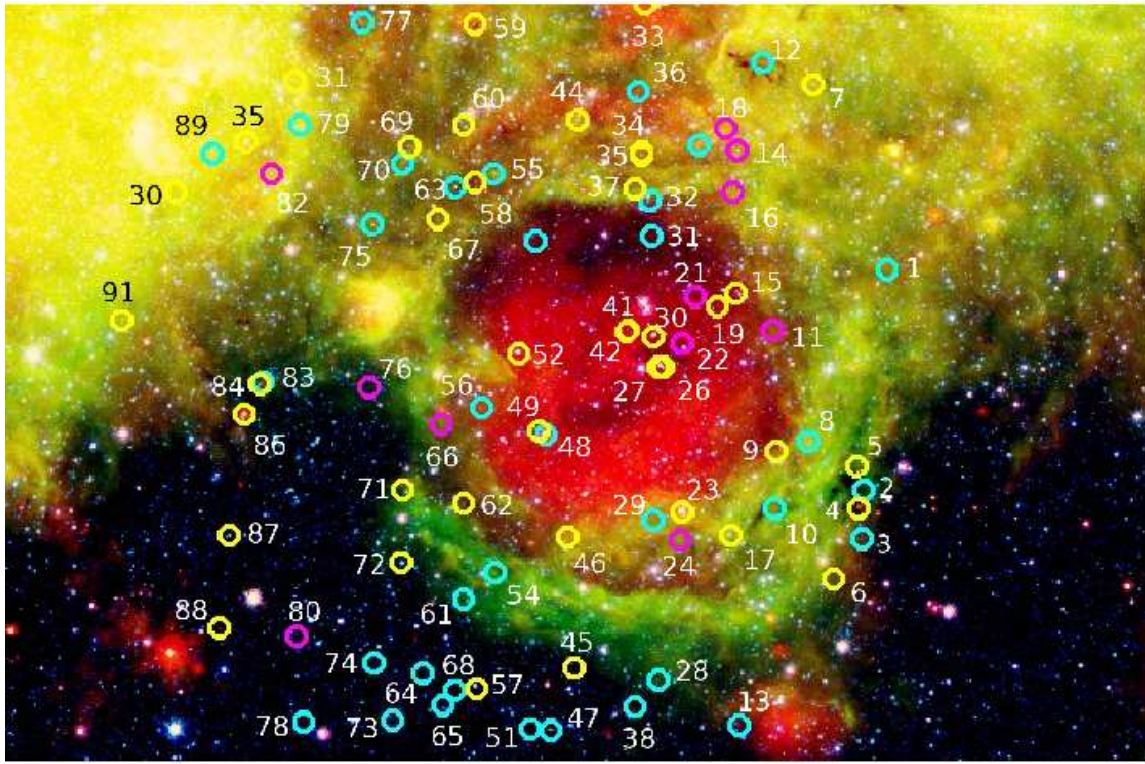}
\end{figure}

All stars shown in Fig.~\ref{cn108ysomap} have been analyzed using the model
fitter of Robitaille et al. (2007) to identify candidate ionizing stars and
YSOs associated with CN108. The locations of candidate YSOs are shown in Fig.~\ref{cn108ysomap} and properties given in Table \ref{ysotable}. The projected YSO
density is lower inside the 8 $\mu$m shell and higher outside the shell to the
lower and upper left. There do not appear to be preferred areas of
concentrated star-formation along the rim, in contrast to CN138. There also
appears to be significant star-formation beyond the 8 $\mu$m PAH emission.
This characteristic was observed by Zavagno et al.  (2007) in their study of
RCW 120. They suggested that a HII region that leaks UV photons (h$\nu$ $>$
13.6 eV) may be carving small scale tunnels through the PDR, inducing
small-scale star-formation far from the ionization front. If such a process is occurring around CN108, we do not observe evidence of
small-scale radio continuum emission within the PDR, as they did using
H$\alpha$. It is possible, however, that such small-scale structure is below the radio-continuum sensitivity limit. 

\begin{figure}
\caption{CN108: Candidate ionizing stars superimposed on 24 $\mu$m (red), 8 $\mu$m (green) and 4.5 $\mu$m (blue) emission. Not all these sources are ionizing the region, but at least two are likely important, one from 3 and 6 and at least one from 1, 2, 4 and 5.}
\label{cn108ionizingmap}
\includegraphics[angle=0,width=6in]{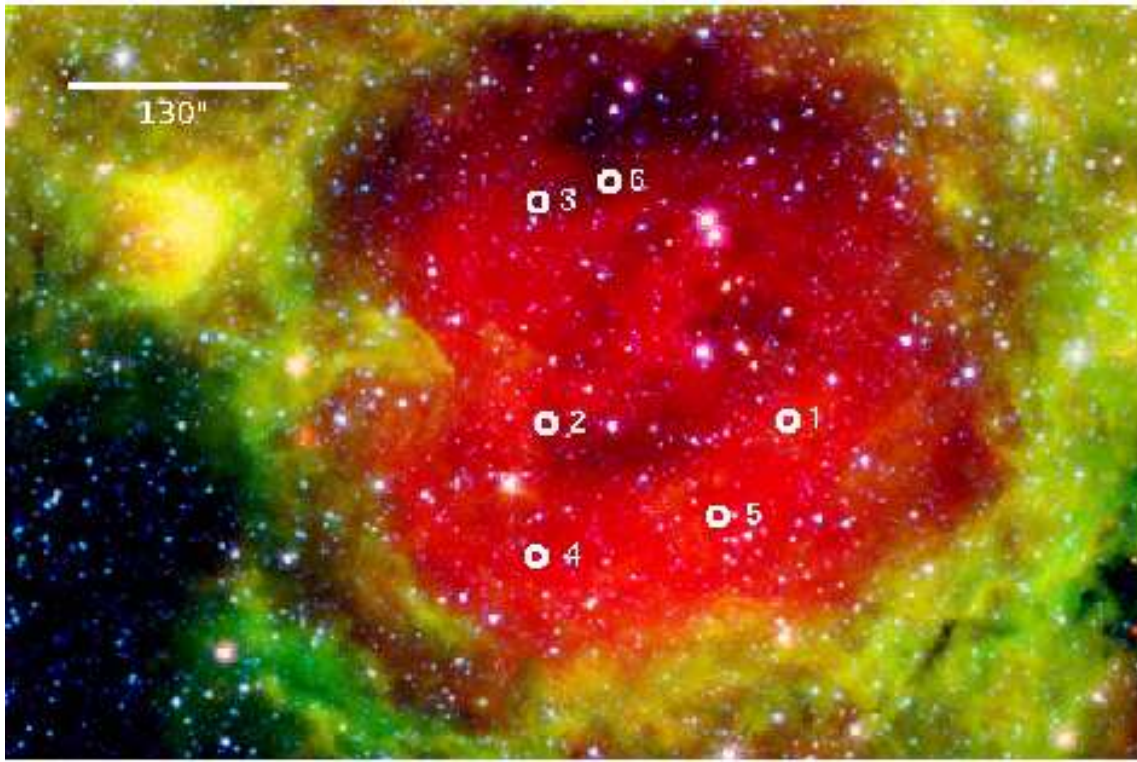}
\end{figure}

Using the method of Watson et al. (2008), we have identified 6 sources whose broadband SEDs are consistent with O-type stars at
the distance of CN108 (see Table~\ref{obtable} and Fig.~\ref{cn108ionizingmap}). The sources are grouped in two clumps, one
near the center of the upper half of CN108, the other along the 24 $\mu$m
emission in the lower half of CN108.  This split and the 8 $\mu$m emission dip
toward the center at about $l$=8.18$^\circ$, $b$=-0.48$^\circ$ implies that
there may be two or three sources creating this bubble.  Considering that the
radio continuum emission implies a single O8 star, however, some of these candidate ionizing stars are likely foreground stars. Unfortunately, without more
constraints, we are unable to further isolate the ionizing stars. Because there are
likely multiple exciting sources, the shell morphology observed at 24 $\mu$m is
probably not produced by a wind-blown cavity (as in N49, Watson et al. 2007),
but rather hot dust centered on each ionizing star. As further evidence that suggests multiple ionizing sources, the 8 $\mu$m emission bubble is scalloped and has multiple centers of curvature. For example, the 8 $\mu$m emission at the upper-left in Fig. \ref{cn108ionizingmap} curves around source 3 whereas the emission at the lower-left curves around source 4.

\subsection{CS57}
\begin{figure}
\caption{Top: CS57 shown in 24 $\mu$m (red), 8 $\mu$m (green) and 4.5 $\mu$m
  (blue). 6 cm in contours. No 20 cm emission was detected above 2 mJy/beam. Note that the 24
  $\mu$m emission is saturated at the bottom of the shell. Bottom: The slice
  is at $b$=-0.14$^\circ$ showing 24 $\mu$m (dashed), and 8.0 $\mu$m (solid, x 5) emission.}
\label{cs57_3color}
\includegraphics[width=5in]{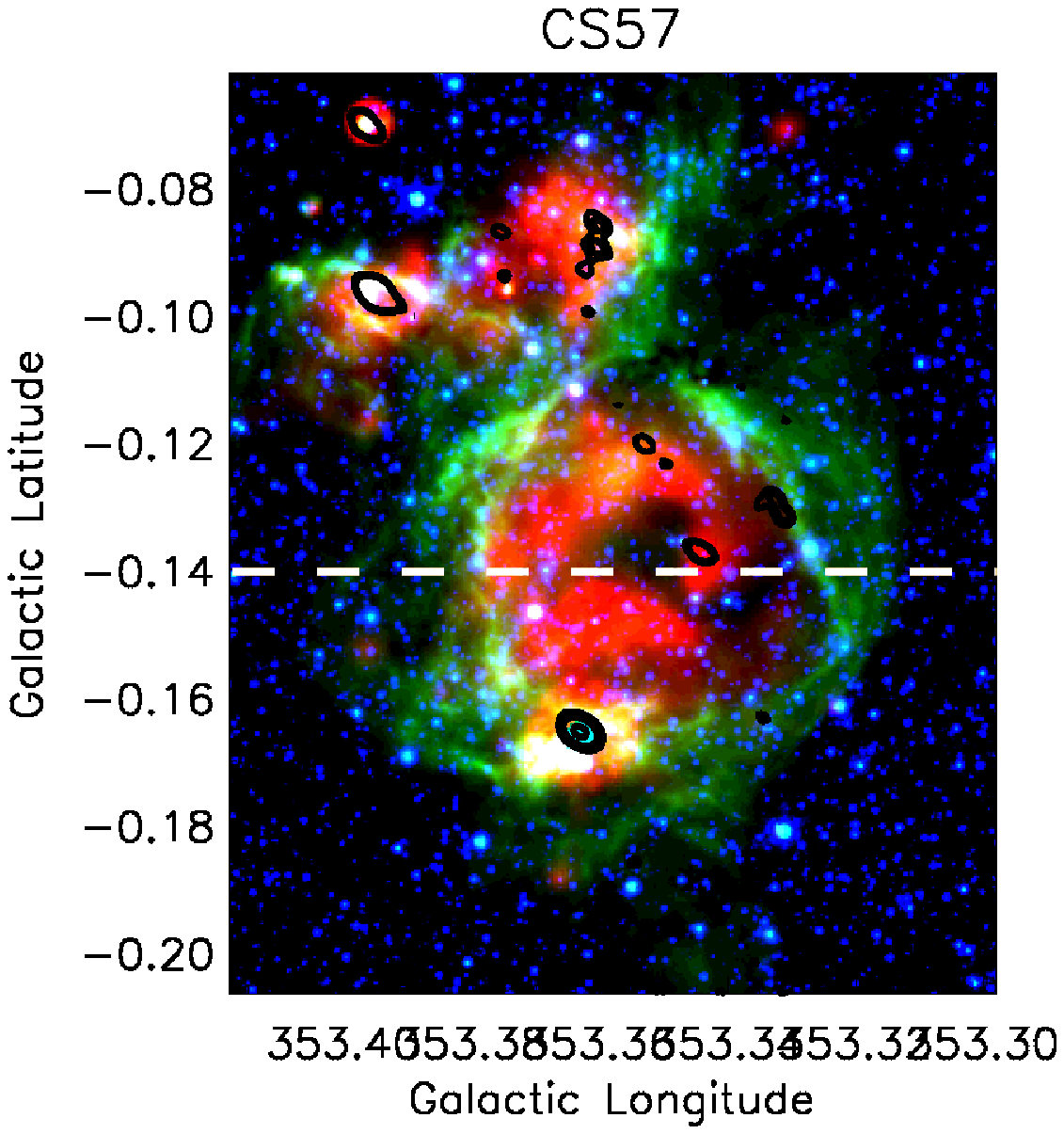}
\includegraphics[width=5in]{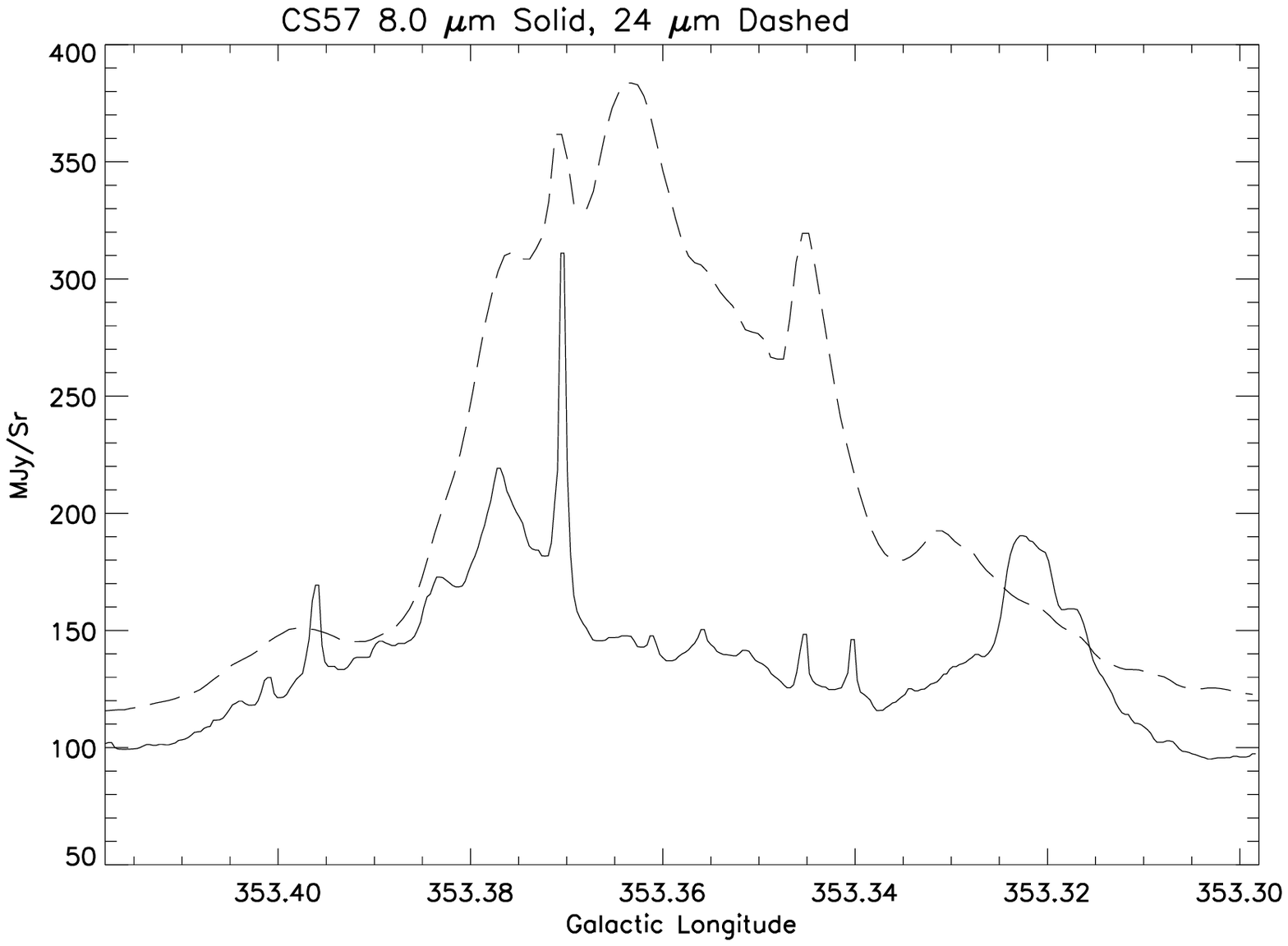}
\end{figure}

CS57 has a shell morphology at 8 $\mu$m surrounding a smaller shell of 24
$\mu$m emission (see Figure~\ref{cs57_3color}). 20 cm emission was not detected toward the center of CS57 coincident with the 24 $\mu$m emission at
a surface brightness level $\geq$ 2 mJy/beam with a beamsize of
7''$\times$4'' (Helfand et al. 2006). 6 cm emission, however, was detected. Several 6 cm (contours in Figure~\ref{cs57_3color}) sources are observed along the 8 $\mu$m shell
and outside the bubble. No radio continuum flux is detected
toward the interior of the shell. Its velocity is -58 km s$^{-1}$, corresponding to a near
kinematic distance of 6.2 kpc. We measure an average inner radius at 8 $\mu$m of 81''
(4.9 pc) and an average outer radius of 130'' (7.6 pc). Using the inner radius, the
near kinematic distance and the limitation on 20 cm emission, we calculate an
upper limit of N$_{Ly}<$ 3.4$\times$10$^{47}$ ionizing photons s$^{-1}$,
equivalent to a star cooler than O9.5V (MSH05). The radius of the 24 $\mu$m emission (measured
to the brightest intensity ridge) is 45'' (2.7 pc).  

The brightest 6 cm radio continuum source (at the bottom of the shell in
Figure~\ref{cs57_3color}) has an integrated flux density of 350 mJy; 24 $\mu$m emission is saturated and point-like. The small, bright flux at 8 $\mu$m ($\sim$ 10 Jy) and 24 $\mu$m ($>$ 2 Jy) and $\frac{8 \mu m}{24 \mu m}$ ratio ($\lesssim$ 5) suggests this source may be an AGB star (see Robitaille et al. 2008), in which case it would likely be unrelated to the bubble. However, the 6 cm emission is consistent with a UCHII region. In such a scenario, the high $\frac{8 \mu m}{24 \mu m}$ ratio could be caused by bright PAH emission in the 8 $\mu$m band. Assuming optically-thin, free-free emission, this flux density implies a UV flux of 9.5 $\times$ 10$^{46}$ photons s$^{-1}$, equivalent to an early B star (MSH05). However, massive stars typically form in regions of high star-formation of all masses, which we do not observe here. Sugitana et al. (1991) argue that intermediate-mass (1.5-6.0 M$_\odot$) star-formation may be caused preferentially by triggering. In summary, we cannot definitively classify this source as either an AGB star or massive YSO.

\begin{figure}
\caption{CS57: Candidate YSOs superimposed on 24 $\mu$m (red), 8 $\mu$m (green) and 4.5 $\mu$m (blue) emission. Stage I sources are shown in yellow, stage II in cyan and stage III in green.}
\label{cs57yso}
\includegraphics[angle=0,width=6in]{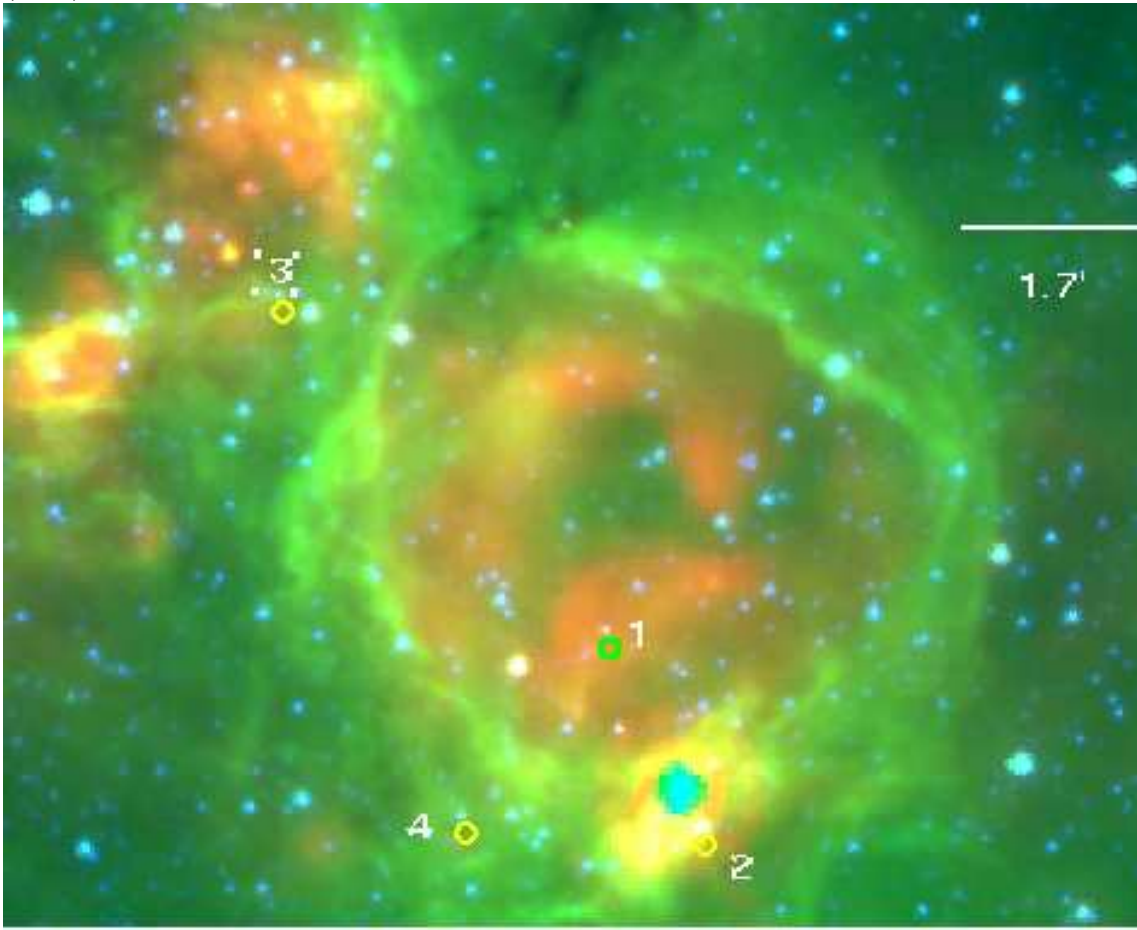}
\end{figure}

We have identified 4 sources as probable YSOs associated with CS57 (see Figure~\ref{cs57yso} for locations and Table \ref{ysotable} for properties). All the YSOs have M$<$10 M$_\odot$ and lie either along the 8 $\mu$m shell or the 24 $\mu$m shell. The low number of YSO candidates is consistent with CS57 representing the low-power end of the bubble population.

We analyzed the GPSC sources to identify the exciting star(s) of CS57. The only
two candidates were O6V stars, however. These spectral types are strongly inconsistent with the absence of radio continuum emission from the bubble center and thus are likely low-mass, foreground stars.
Identifying the mid-to-early B star responsible for this bubble is difficult
because it is easy to confuse a mid-to-early B star at the bubble's location with a
cool, foreground star.  

\section{Analysis: Triggered Star-Formation}
\begin{figure}
\epsscale{.55}
\caption{Numerical simulation of the expansion of an HII region, model S19 modified by original authors (Hosokawa, priv. comm.) expanding into a dense
  ISM (3 $\times$ 10 $^3$ cm$^{-3}$). Model S19 is Fig. 11 from Hosokawa \& Inutsuka (2006). The ionizing
  star has a mass M$_*$ = 19 M$_\odot$. The simulations were 1D and incorporated radiative heating and cooling, photo- and collisional ionization,
photodissociation and recombination.} 
\label{hif11}
\plotone{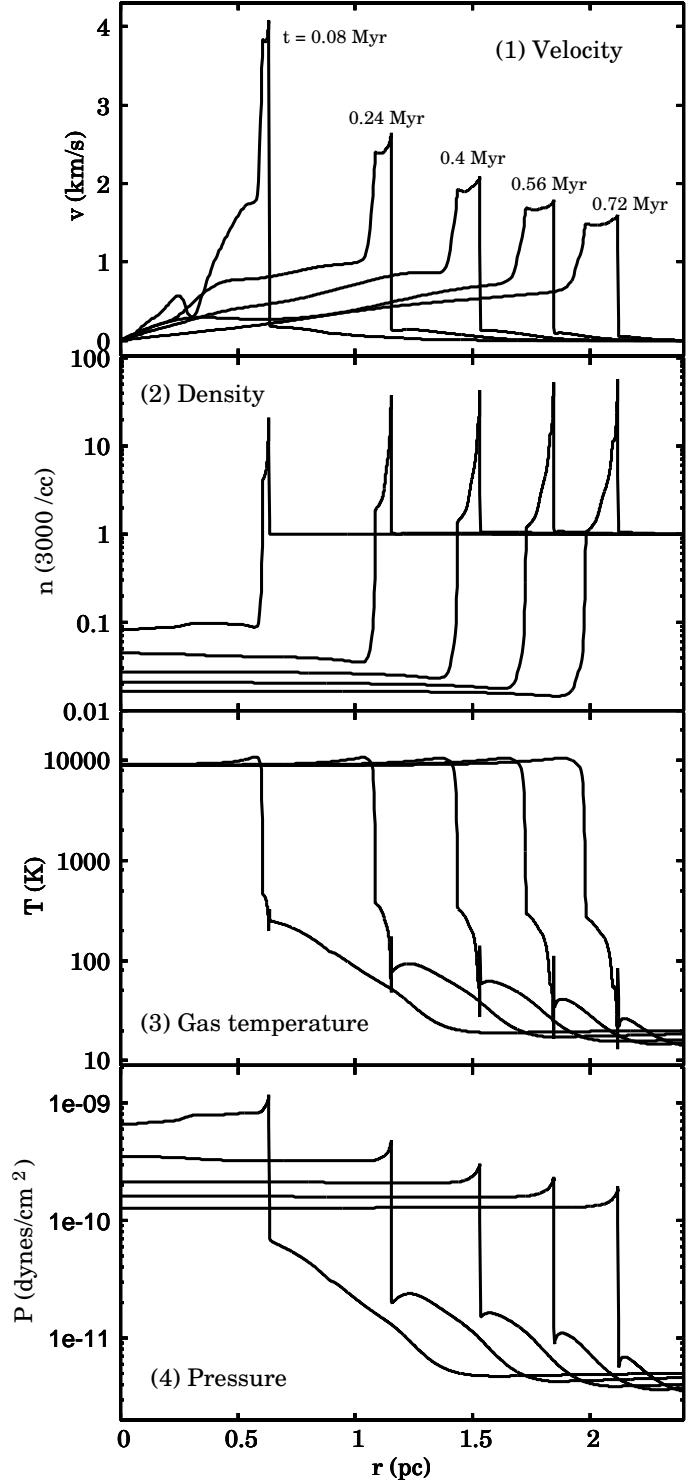}
\end{figure}

We now discuss the likelihood of triggered star-formation in each region. Hosokawa \& Inutsuka (2006) (hereafter HI06) present numerical
simulations of expanding HII regions and their associated PDRs. Their
models concentrate on incorporating the different physics in the HII
region and surrounding swept-up shell. Specifically, they treat the
UV- and FUV- radiative transfer and cooling, photo- and collisional ionization,
photodissociation and recombination in the HII region and PDR region (see their Table 2 and
3 for a summary of the dominant cooling lines in the HII region and PDR). They were principally interested in modeling the
progression of the ionization front, dissociation front and shock
front due to an overpressured HII region surrounding a single massive
star. Their models do not include stellar winds or cooling in the HII
region due to dust emission. We do not expect stellar winds to
dominate the late-O and early-B stars we are studying here. Cooling
due to dust emission may be important, however, but has not been
well-studied. HI06 also calculate the time required for swept-up
ambient gas to become gravitationally unstable to collapse, possibly
resulting in triggered star-formation (i.e. the collect-and-collapse
mechanism). They present 5 models of different central stellar masses
(11.7 M$_\odot$ - 101.3 M$_\odot$) and ambient densities (10$^2$
cm$^{-3}$ - 10$^4$ cm$^{-3}$). The authors have provided two new models with parameters adjusted so that the observational predictions match the observations presented here. We use a model similar to their model S19, which has a central mass of 19 M$_\odot$. The authors have adjusted the ambient density to 3 $\times$ 10$^3$ cm$^{-3}$. The velocity, density, gas
temperature and pressure distributions at different epochs for this
model are shown in Figure~\ref{hif11}.

\begin{figure}
\epsscale{1}
\caption{Numerical simulation of an HII region (model S19 modified by Hosokawa, priv. comm.) expanding into a dense
  ISM (3 $\times$ 10 $^3$ cm$^{-3}$). The dynamics of the original model are shown in Fig. 14 from Hosokawa \& Inutsuka (2006). The ionizing
  star has a mass M$_*$ = 19 M$_\odot$.  Top: The position of the shock front (SF), dissociation front (DF) and ionization front (IF) as the HII region expands due to internal overpressure. We use this model to estimate the age
  of CN138 by matching the observed size of the 8 $\mu$m shell (see dashed line). We also use this model to conclude that the presence of triggered star-formation along the shell is consistent with the size of the bubble.}
\label{hif14}
\plotone{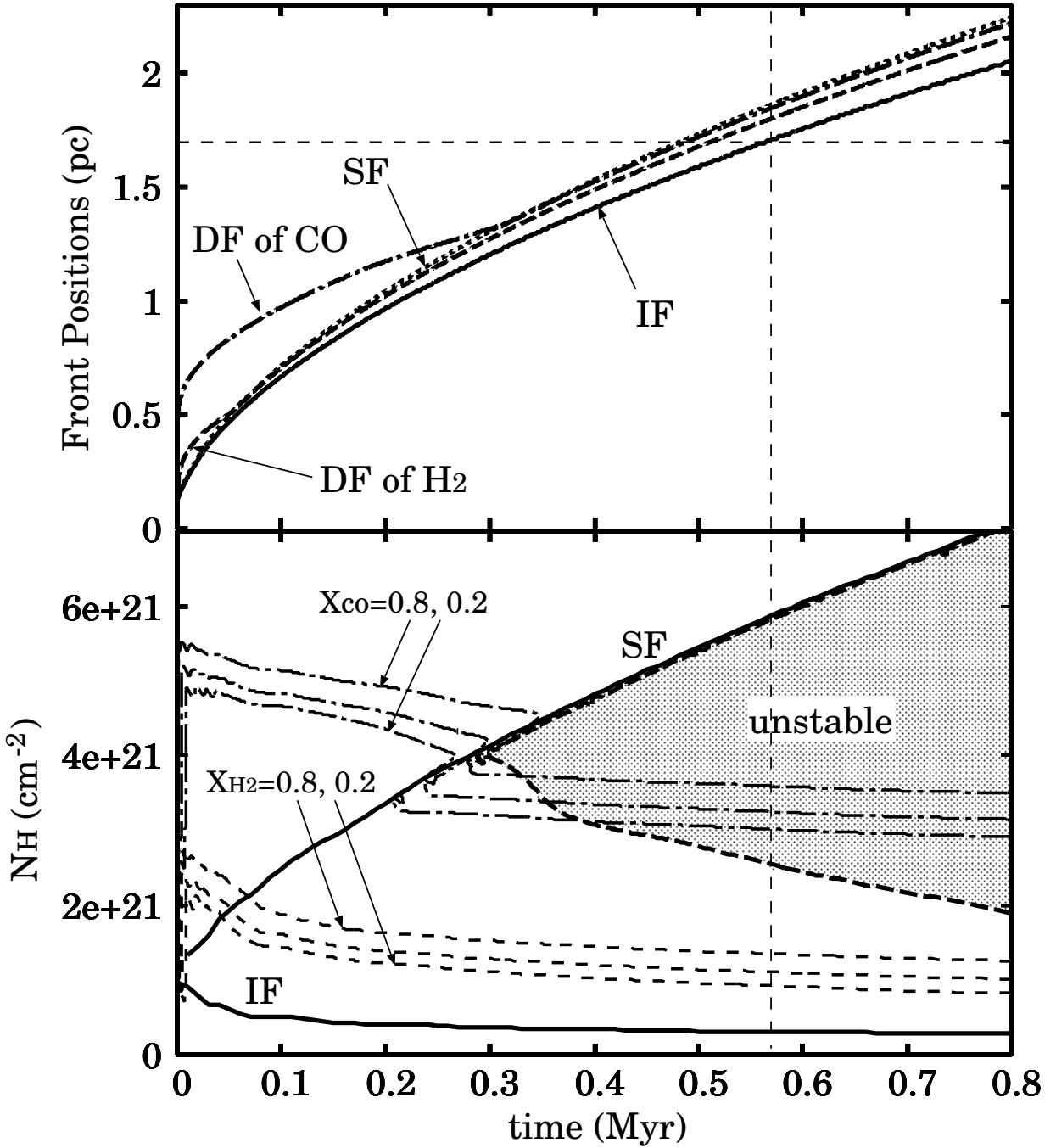}
\end{figure}

If we equate the 8 $\mu$m emission inner radius observed here with the position of the H$_2$ dissociation front, shock front and ionization front in this model (which are nearly coincident), we can estimate the dynamical age of the bubble. Figure \ref{hif14} (top panel) shows the growth of these three fronts' radii over time for the modified version of model S19 described above. Based on this assumption, CN138 appears quite young, 0.5-0.6 Myrs. HI06 also determine when the shell density would become gravitationally unstable to collapse, implying when triggered star-formation could be expected to begin (see Figure \ref{hif14}, bottom panel). In their modified model, gravitational collapse begins at t=0.3 Myrs. Thus, we conclude that the size of CN138 is consistent with the presence of triggered star-formation. These results, however, are strongly dependent on the assumed ambient density. If we use the published model S19 with an ambient density of 10$^3$ cm$^{-3}$, we would conclude that the bubble age is 0.15-0.45 Myrs and gravitational collapse begins at t=0.5 Myrs, implying that triggered star-formation has not started.

Since CN108 appears to involve several bubbles overlapping, its
expansion is significantly more complicated than either CN138 or
CS57. No model by HI06 includes multiple sources, so we cannot
estimate its dynamical age. CS57, on the other hand, is difficult to
interpret because we do not know the spectral type of the ionizing
star. We use a modified version lowest-mass model presented by HI06 (model S12,
M$_*$ = 11 M$_\odot$, n$_{ambient}$ = 100 cm$^{-3}$, Hosokawa, priv. comm.). The measured radius (4.9-7.6 pc) is consistent with an age of 3.5-7 Myrs. At this stage, Hosokawa (priv. comm.) predicts that the shell will be gravitationally unstable. This prediction is consistent with the YSOs we detect along the shell being triggered by the expanding shell.

\section{Conclusions}
We have analyzed three late-O/early-B-powered bubbles from the catalogue of Churchwell et al. (2007). Our conclusions are as follows:\\
$\bullet$ Similar to the higher energy bubbles analyzed previously, each bubble shows the same basic structure, a PDR surrounding hot
dust and, in two of three sources, ionized gas.\\
$\bullet$ Potential triggered star-formation by the collect-and-collapse mechanism has been identified in two
bubbles, CN138 and CS57.\\
$\bullet$ Candidate ionizing stars are identified in CN138 and CN108. Based on
spectral types implied by their SEDs and radio continuum emission, the bubbles do not appear to be
wind-dominated. CN138 appears to be driven by one or two stars that are off-center with spectral types O8.5 and O9. CN108, on the other hand, appears to be driven by two or three hot stars with spectral type between O6 and O9.5.\\
$\bullet$ The age of two bubbles are approximated through comparison with modified versions of the
numerical models of HI06. The age of both CS57 and CN138 are consistent with the presence of
the identified YSOs being triggered by the bubbles' expansion.

\acknowledgements
We would like to acknowledge support for this work by NASA contracts 1289406 and 1275394. We would also like to acknowledge the helpful comments of T. Hosokawa, and especially for providing new simulations for comparison with CN138 and CS57.


\begin{deluxetable}{lcrrrrrrrrrrrr}
 \tabletypesize{\tiny}
 \rotate
\tablecaption{Model parameters for candidate YSOs}
\tablehead{
  \multicolumn{2}{c}{ } & \multicolumn{2}{c}{$M_{\star}$ (M$_{\sun}$)} & \multicolumn{2}{c}{$L_{\rm TOT}$ (L$_{\sun})$} & \multicolumn{2}{c}{$\dot{M}_{\rm env}$ (M$_{\sun}$ yr$^{-1}$)}\\ 
  \colhead{ID} & \colhead{Name (G$l+b$)} & \colhead{min} & \colhead{max} & \colhead{min} & \colhead{max}  & \colhead{min} & \colhead{max} &Stage\\
}
\startdata
CN138-1         &G9.7813-0.7732  &0.7    &4.2    &5      &337                 &0               &6.13E$-$6       &III\\
CN138-2         &G9.7861-0.6814  &0.1    &18.4   &2      &13880               &0               &3.06E$-$3       &I  \\
CN138-3         &G9.7965-0.7729  &0.3    &7.1    &3      &932                 &0               &6.18E$-$4       &I  \\
CN138-4         &G9.8154-0.6861  &0.2    &6.2    &2      &1019                &0               &2.35E$-$4       &II \\
CN138-5         &G9.8261-0.7127  &0.3    &6.3    &5      &1100                &0               &5.38E$-$4       &I  \\
CN138-6         &G9.8265-0.7172  &3.6    &14.5   &140    &7305                &2.25E$-$5       &1.88E$-$3       &\nodata\\
CN138-7         &G9.8268-0.7143  &0.7    &10.5   &8      &2306                &0               &2.10E$-$3       &I  \\
CN138-8         &G9.8291-0.6707  &1.8    &4.4    &28     &127                 &1.03E$-$8       &5.63E$-$5       &III\\
CN138-9         &G9.8292-0.6478  &0.5    &5.0    &2      &469                 &0               &1.88E$-$5       &II \\
CN138-10        &G9.8350-0.6358  &5.3    &10.9   &544    &8256                &0               &1.32E$-$3       &II \\
CN138-11        &G9.8366-0.6479  &7.9    &7.9    &745    &745                 &1.56E$-$4       &1.56E$-$4       &I  \\
CN138-12        &G9.8486-0.7176  &1.0    &6.5    &24     &331                 &4.18E$-$7       &1.44E$-$3       &I  \\
CN138-13        &G9.8523-0.7226  &2.7    &7.3    &49     &1942                &0               &0               &III\\
CN138-14        &G9.8529-0.7285  &0.2    &7.3    &9      &1995                &0               &9.36E$-$4       &II \\
CN138-15        &G9.8705-0.7731  &0.4    &5.4    &5      &321                 &0               &6.18E$-$4       &I  \\
CN138-16        &G9.8736-0.7553  &1.6    &22.5   &48     &61250               &1.35E$-$5       &4.68E$-$3       &I  \\
CN138-17        &G9.8803-0.7506  &0.2    &20.0   &13     &35300               &0               &1.86E$-$3       &I  \\
CN138-18        &G9.8849-0.6539  &9.3    &19.2   &5882   &41650               &0               &0               &II \\
CN138-19        &G9.8991-0.7493  &0.2    &13.4   &9      &7831                &0               &2.33E$-$3       &II \\
CN138-20        &G9.9067-0.6962  &0.4    &7.0    &5      &583                 &0               &6.18E$-$4       &III\\
\hline
CN108-1         &G7.9828-0.4212  &0.3    &4.0    &2      &249                 &0               &1.22E$-$4       &II\\
CN108-2         &G7.9954-0.5735  &0.6    &3.9    &4      &249                 &0               &3.43E$-$5       &II\\
CN108-3         &G7.9960-0.5638  &0.6    &4.5    &4      &334                 &0               &2.35E$-$4       &II\\
CN108-4         &G7.9980-0.5471  &0.6    &5.6    &28     &697                 &0               &1.63E$-$3       &I \\
CN108-5         &G7.9985-0.5252  &0.3    &8.4    &2      &1159                &0               &1.78E$-$3       &I \\
CN108-6         &G8.0115-0.5845  &0.4    &4.1    &3      &234                 &0               &1.47E$-$4       &I \\
CN108-7         &G8.0219-0.3233  &0.4    &10.6   &11     &2080                &0               &1.06E$-$3       &I \\
CN108-8         &G8.0243-0.5117  &0.2    &8.8    &1      &1159                &0               &2.81E$-$3       &II\\
CN108-9         &G8.0416-0.5169  &0.2    &4.6    &2      &361                 &0               &1.80E$-$4       &I\\
CN108-10        &G8.0428-0.5472  &0.4    &4.1    &3      &249                 &0               &8.85E$-$5       &II\\
CN108-11        &G8.0430-0.4532  &0.3    &4.1    &2      &249                 &0               &1.80E$-$4       &III\\
CN108-12        &G8.0483-0.3122  &0.8    &5.1    &6      &487                 &0               &2.35E$-$4       &II \\
CN108-13        &G8.0612-0.6618  &3.9    &4.0    &52     &53                  &3.43E$-$7       &3.66E$-$7       &II \\
CN108-14        &G8.0620-0.3586  &3.7    &11.6   &208    &9883                &0               &0               &III\\
CN108-15        &G8.0625-0.4334  &1.1    &5.1    &13     &100                 &2.31E$-$7       &2.40E$-$4       &I  \\
CN108-16        &G8.0642-0.3804  &2.3    &3.5    &28     &121                 &0               &0               &III\\
CN108-17        &G8.0653-0.5617  &4.8    &8.7    &187    &1217                &2.53E$-$5       &2.37E$-$3       &I  \\
CN108-18        &G8.0686-0.3466  &0.7    &3.5    &4      &119                 &0               &5.04E$-$7       &III\\
CN108-19        &G8.0725-0.4408  &0.1    &7.9    &2      &2734                &0               &5.41E$-$4       &I  \\
CN108-20        &G8.0823-0.3555  &0.5    &5.0    &4      &469                 &0               &1.22E$-$4       &II \\
CN108-21        &G8.0845-0.4357  &2.3    &3.5    &28     &121                 &0               &0               &III\\
CN108-22        &G8.0908-0.4606  &0.2    &4.6    &1      &361                 &0               &2.37E$-$4       &III\\
CN108-23        &G8.0911-0.5492  &8.0    &16.2   &2714   &17220               &0               &6.90E$-$3       &I\\
CN108-24        &G8.0924-0.5646  &0.4    &4.5    &5      &283                 &0               &4.37E$-$4       &III\\
CN108-25        &G8.1012-0.2742  &2.9    &11.8   &94     &10220               &0               &1.07E$-$3       &I\\
CN108-26        &G8.1014-0.4731  &2.3    &10.7   &81     &4922                &0               &7.44E$-$4       &I\\
CN108-27        &G8.1040-0.4732  &0.7    &7.6    &12     &602                 &6.94E$-$9       &4.69E$-$4       &I\\
CN108-28        &G8.1042-0.6382  &0.6    &3.9    &7      &55                  &1.93E$-$9       &1.71E$-$4       &II\\
CN108-29        &G8.1063-0.5533  &2.0    &4.5    &19     &323                 &0               &0               &II\\
CN108-30        &G8.1066-0.4568  &0.2    &8.5    &9      &3554                &0               &9.24E$-$4       &I \\
CN108-31        &G8.1076-0.4037  &0.6    &4.4    &6      &148                 &0               &4.13E$-$5       &II\\
CN108-32        &G8.1085-0.3852  &0.9    &5.1    &7      &499                 &0               &1.01E$-$4       &II\\
CN108-33        &G8.1110-0.2798  &2.0    &4.4    &38     &127                 &1.88E$-$6       &1.97E$-$4       &I \\
CN108-34        &G8.1129-0.3593  &0.4    &8.8    &2      &1211                &0               &2.15E$-$3       &I \\
CN108-35        &G8.1129-0.3610  &1.4    &9.6    &44     &5151                &0               &2.33E$-$3       &I \\
CN108-36        &G8.1143-0.3268  &0.7    &6.2    &8      &1543                &0               &2.04E$-$4       &II\\
CN108-37        &G8.1161-0.3783  &0.6    &5.1    &7      &250                 &0               &4.37E$-$4       &I \\
CN108-38        &G8.1161-0.6525  &0.5    &3.9    &3      &249                 &0               &1.33E$-$4       &II\\
CN108-39        &G8.1176-0.2717  &1.0    &5.2    &12     &383                 &0               &2.18E$-$4       &II\\
CN108-40        &G8.1184-0.2634  &0.6    &5.5    &4      &469                 &0               &4.24E$-$5       &II\\
CN108-41        &G8.1201-0.4529  &1.1    &7.5    &27     &1690                &3.98E$-$6       &1.79E$-$3       &I \\
CN108-42        &G8.1203-0.4539  &1.6    &7.5    &38     &1017                &1.20E$-$5       &1.79E$-$3       &I \\
CN108-43        &G8.1365-0.2664  &0.7    &8.0    &11     &1224                &0               &1.79E$-$3       &I \\
CN108-44        &G8.1468-0.3420  &0.2    &4.6    &1      &361                 &0               &2.50E$-$4       &I \\
CN108-45        &G8.1483-0.6319  &0.5    &4.2    &3      &249                 &0               &4.13E$-$5       &I \\
CN108-46        &G8.1522-0.5630  &0.3    &4.0    &2      &213                 &0               &8.85E$-$5       &I \\
CN108-47        &G8.1609-0.6646  &4.6    &8.4    &383    &3447                &0               &0               &II\\
CN108-48        &G8.1634-0.5083  &11.3   &23.7   &9714   &70880               &0               &0               &II\\
CN108-49        &G8.1667-0.5062  &0.7    &7.8    &10     &631                 &8.82E$-$8       &9.41E$-$4       &I\\
CN108-50        &G8.1684-0.4060  &0.5    &4.6    &4      &361                 &0               &1.71E$-$4       &II\\
CN108-51        &G8.1716-0.6643  &0.5    &3.8    &4      &178                 &0               &1.22E$-$4       &II\\
CN108-52        &G8.1774-0.4662  &0.4    &8.7    &5      &1217                &0               &2.37E$-$3       &I \\
CN108-53        &G8.1832-0.2707  &0.4    &4.6    &4      &361                 &0               &2.50E$-$4       &I \\
CN108-54        &G8.1903-0.5817  &0.5    &4.2    &3      &337                 &0               &7.78E$-$5       &II\\
CN108-55        &G8.1906-0.3702  &0.2    &7.5    &4      &2150                &0               &4.67E$-$4       &II\\
CN108-56        &G8.1973-0.4937  &0.5    &4.2    &5      &161                 &0               &1.02E$-$4       &II\\
CN108-57        &G8.2000-0.6423  &2.0    &7.5    &40     &510                 &3.31E$-$6       &6.73E$-$4       &I \\
CN108-58        &G8.2003-0.3751  &1.5    &4.7    &19     &144                 &0               &1.37E$-$4       &I \\
CN108-59        &G8.2010-0.2917  &0.9    &6.8    &32     &464                 &8.10E$-$8       &8.65E$-$4       &I \\
CN108-60        &G8.2066-0.3452  &2.1    &8.0    &40     &1017                &1.20E$-$5       &1.79E$-$3       &I \\
CN108-61        &G8.2067-0.5951  &0.4    &4.6    &2      &361                 &0               &6.65E$-$5       &II\\
CN108-62        &G8.2070-0.5448  &0.2    &4.3    &1      &261                 &0               &1.29E$-$4       &I \\
CN108-63        &G8.2109-0.3777  &3.2    &6.7    &129    &1376                &0               &0               &II\\
CN108-64        &G8.2113-0.6432  &0.1    &4.1    &1      &249                 &0               &1.80E$-$4       &II\\
CN108-65        &G8.2178-0.6518  &0.8    &4.5    &8      &145                 &0               &1.02E$-$4       &II\\
CN108-66        &G8.2188-0.5030  &0.3    &4.0    &2      &249                 &0               &1.80E$-$4       &III\\
CN108-67        &G8.2205-0.3947  &0.6    &5.1    &16     &509                 &0               &1.27E$-$3       &I  \\
CN108-68        &G8.2285-0.6343  &0.6    &3.9    &5      &74                  &0               &4.13E$-$5       &II \\
CN108-69        &G8.2352-0.3560  &0.9    &7.3    &7      &1942                &0               &4.67E$-$4       &I  \\
CN108-70        &G8.2386-0.3649  &0.6    &6.2    &6      &1038                &0               &4.13E$-$5       &II \\
CN108-71        &G8.2389-0.5381  &0.2    &10.3   &1      &1863                &0               &2.37E$-$3       &I  \\
CN108-72        &G8.2399-0.5755  &0.5    &6.2    &7      &997                 &0               &3.76E$-$4       &I  \\
CN108-73        &G8.2439-0.6591  &3.9    &4.0    &52     &53                  &3.43E$-$7       &3.66E$-$7       &II \\
CN108-74        &G8.2537-0.6294  &1.7    &3.8    &10     &157                 &0               &1.88E$-$5       &II \\
CN108-75        &G8.2547-0.3976  &2.7    &5.7    &55     &733                 &0               &3.25E$-$7       &II \\
CN108-76        &G8.2568-0.4835  &0.1    &4.6    &1      &361                 &0               &1.80E$-$4       &III\\
CN108-77        &G8.2602-0.2906  &2.9    &5.9    &63     &843                 &0               &2.07E$-$7       &II \\
CN108-78        &G8.2913-0.6607  &0.7    &7.3    &5      &1942                &0               &2.51E$-$6       &II \\
CN108-79        &G8.2928-0.3444  &0.7    &4.8    &11     &383                 &0               &4.99E$-$4       &II \\
CN108-80        &G8.2948-0.6148  &1.0    &7.3    &11     &1942                &0               &4.67E$-$4       &III\\
CN108-81        &G8.2956-0.3215  &1.5    &5.2    &19     &499                 &0               &2.40E$-$4       &I  \\
CN108-82        &G8.3082-0.3702  &1.3    &4.7    &16     &383                 &0               &1.37E$-$4       &III\\
CN108-83        &G8.3127-0.4811  &2.6    &5.5    &32     &665                 &0               &1.62E$-$6       &II\\
CN108-84        &G8.3144-0.4821  &0.7    &8.8    &17     &1847                &0               &9.76E$-$4       &I\\
CN108-85        &G8.3217-0.3540  &2.4    &9.8    &186    &5808                &0               &2.63E$-$3       &I\\
CN108-86        &G8.3224-0.4976  &17.3   &49.6   &37170  &376600              &1.11E$-$3       &4.94E$-$3       &I\\
CN108-87        &G8.3301-0.5619  &4.2    &7.3    &194    &1717                &0               &7.65E$-$4       &I\\
CN108-88        &G8.3357-0.6110  &1.2    &6.6    &33     &1324                &0               &1.33E$-$3       &I\\
CN108-89        &G8.3393-0.3598  &3.2    &6.4    &119    &1132                &0               &2.23E$-$7       &II\\
CN108-90        &G8.3585-0.3805  &0.2    &7.7    &1      &548                 &0               &1.59E$-$3       &I\\
CN108-91        &G8.3868-0.4480  &0.6    &7.7    &9      &566                 &0               &9.41E$-$4       &I\\
\hline
CS57-1  &G353.3612-0.1481        &1.0    &11.6   &110    &10020               &0               &1.62E$-$3       &I\\
CS57-2  &G353.3629-0.1703        &6.0    &20.0   &993    &46230               &0               &4.03E$-$3       &II\\
CS57-3  &G353.3727-0.1019        &2.3    &11.5   &85     &5111                &0               &2.74E$-$3       &II\\
CS57-4  &G353.3833-0.1570        &3.7    &11.6   &152    &9883                &0               &1.75E$-$3       &II\\
\enddata
\label{ysotable}
\end{deluxetable}

\begin{table}
\caption{Ionizing Star Candidates}
\label{obtable}
\begin{tabular}{rrrr}
ID &Name &Spectral type \\
\hline
ICN138-1 &G9.8420-00.7134 &O8.5V \\
ICN138-2 &G9.8421-00.7127 &O9.0V \\
\hline
ICN108-1        &G8.0903-00.4912        &O7.0V\\
ICN108-2        &G8.1090-00.5168        &O9.5V\\
ICN108-3        &G8.1375-00.4282        &O9.5V\\
ICN108-4        &G8.1541-00.4920        &O6.0V\\
ICN108-5        &G8.1565-00.4337        &O7.5V\\
ICN108-6        &G8.1566-00.5274        &O7.0V\\
\hline
\end{tabular}
\end{table}


\end{document}